\documentclass[twocolumn,showpacs,preprintnumbers]{revtex4}

\usepackage{graphicx}
\usepackage{dcolumn}
\usepackage{bm}
\usepackage{amssymb}
\usepackage{amsmath}
\usepackage{epsfig}

\def\lapp{\mathrel{\rlap{\raise.5ex\hbox{$<$}}
                    {\lower.5ex\hbox{$\sim$}}}}
\def\gapp{\mathrel{\rlap{\raise.5ex\hbox{$>$}}
                    {\lower.5ex\hbox{$\sim$}}}}

\def\bc {\begin{center}}
\def\ec {\end{center}}
\def\be {\begin{equation}}
\def\bea {\begin{eqnarray}}
\def\ee {\end{equation}}
\def\eea {\end{eqnarray}}

\def\dam{{(\Delta m_{31}^2)^m}}
\def\da{{(\Delta m_{31}^2)}}

\begin{document}
\vspace*{1cm}
\title {The discrimination of mass hierarchy with atmospheric neutrinos at a magnetized muon detector
}

\author 
{Abhijit Samanta
\footnote{E-mail address: abhijit@hri.res.in}}
\affiliation
{Harish-Chandra Research Institute,
Chhatnag Road,
Jhusi,
Allahabad 211 019,
India}

\date{\today}

\begin{abstract}
\noindent
We have studied the mass hierarchy with atmospheric neutrinos 
considering the muon energy and zenith angle of the event at the magnetized iron calorimeter detector.
For $\chi^2$ analysis 
we have migrated the number of events from  neutrino energy and zenith angle bins 
to muon energy and zenith angle bins using the two-dimensional energy-angle correlated
resolution functions. 
The binning of data is made in two-dimensional grids of $\log_{10} E - L^{0.4}$ plane
to get a better reflection of the oscillation pattern in the $\chi^2$ analysis.
Then the $\chi^2$ is marginalized considering all possible systematic uncertainties of 
the atmospheric neutrino flux and cross section. The effects of the ranges
of  oscillation parameters on the marginalization are also studied. The lower limit of
the range of $\theta_{13}$ for marginalization is found to be very crucial in determining the 
sensitivity of hierarchy for a given $\theta_{13}$. Finally, we show 
that one can discriminate atmospheric neutrino mass hierarchy at $>$90\% C.L. if the lower
limit of $\theta_{13} \ge 5^\circ$.
\end{abstract}



\maketitle
\section{Introduction}
The origin of  masses of the particles  and their interactions have been
successfully described by the standard model (SM). The recent
discovery of neutrino masses and their mixing through neutrino oscillation \cite{Fukuda:1998mi,theory}
opens a new window beyond the SM.
This provides the measurements of mass squared differences 
$\Delta m_{ji}^2 = m_j^2-m_i^2$  and mixing
angles $\theta_{ij}$. 
At present $1 (3) \sigma$ ranges are \cite{Fogli:2008ig}:
$\Delta m_{21}^2 = 7.67^{+0.16}_{-0.19} (^{+0.42}_{-0.53})\times 10^{-5}$eV$^2$, 
$|\Delta m_{31}^2| = 2.39^{+0.11}_{-0.8} (^{+0.42}_{-0.33})\times 10^{-3}$eV$^2$, 
$\sin^2\theta_{12}=0.312^{+0.019}_{-0.018} (^{+0.063}_{-0.049})$,
$\sin^2\theta_{23}=0.466^{+0.058}_{-0.073} (^{+0.178}_{-0.135})$, {and}
$\sin^2\theta_{13}=0.016^{+0.010}_{-0.010} (<0.046)$.
Here $\theta_{ij}$ are the mixing angles in Pontecorvo, Maki, Nakagawa, Sakata 
mixing matrix \cite{mixing}. 
Currently, there is no constraint on the
CP violating phase  $\delta_{\rm CP}$ or on the sign of $\Delta m_{31}^2$.
The neutrino mass ordering is one of the key problems 
to build up the fundamental theory of the particles beyond the SM.
This can be probed through neutrino oscillation. In this paper we have concentrated
on the prospects to resolve the mass hierarchy using atmospheric neutrinos.

The mass hierarchy with atmospheric neutrinos has been studies in 
\cite{Indumathi:2004kd,Gandhi:2007td,Petcov:2005rv} 
with a large magnetized Iron CALorimeter (ICAL) detector, which is being strong considered
for the India-based Neutrino Observatory \cite{Athar:2006yb}. In \cite{Gandhi:2008zs}, the
authors  estimated the sensitivity of a large liquid argon detector. This is nonmagnetized, but can
detect both electron and muon. So,  here the statistics are high compared to ICAL.
However, these studies have dealt with neutrino energy and zenith angle and assumed 
fixed Gaussian resolutions for them separately. 
It is also shown that the confidence level of determining hierarchy changes
drastically with the change in width of resolutions. 
Moreover,  
since all of the particles produced in the neutrino interactions can not be reconstructed, 
the resolutions are not Gaussian in nature. 
There are many neutral
particles in the deep inelastic processes, and the non-Gaussian nature of the resolutions 
increases as one goes to higher energies.
Again, the resolutions are different for neutrino and antineutrino 
(see Fig. 4 of \cite{Samanta:2008af}) since the different quantum numbers are involved
in the scattering matrix elements for neutrino and antineutrino. 
At energies around $1$ GeV, most of the events are quasi-elastic, and the muon carries 
almost all of the
energy of the neutrino. Here, the energy resolution is very good. As the energy increases, 
the deep inelastic process dominates and the energy resolution begins to worsen 
by developing a more prominent non-Gaussian nature.  However, the trend of
angular resolution is quite opposite, its width is very wide at low energy, and it improves
with an increase in energy. So, it is very important to study the sensitivity of a detector
in terms of directly measurable quantities. 

We have studied the neutrino mass hierarchy for the magnetized ICAL detector with atmospheric neutrinos
generating events by NUANCE-v3 \cite{Casper:2002sd} and
considering the muon energy and direction (directly measurable quantities) 
of the event. 
Because of heavy mass of the muon, it looses energy mostly via ionization and atomic excitation during 
its propagation through a medium. Since  ICAL is a tracking detector, it gives a clean single 
track in the detector. The muon energy can be measured from the bending of the track in 
the magnetic field or from the track length in the case of a fully contained event. The direction can be 
measured from the tangent of the track at the vertex.
From the GEANT \cite{geant} simulation of the ICAL detector it is found that the energy and 
angular resolutions of the muons are very
high (4-10\% for energy and 4-12\% for zenith angle) and negligible compared to the resolutions obtained
from the kinematics of the scattering processes. A new method for migration from true neutrino 
energy and zenith angle to muon energy and zenith angle has been introduced in \cite{Samanta:2006sj}
and subsequently used in \cite{Samanta:2008af,Samanta:2008ag,Samanta:2009hd}.

On the other hand, the binning of the data is also an important issue
when one considers the binning of the events in reconstructed energy and zenith angle bins.  
The reasons are the following: The atmospheric neutrino flux changes very rapidly with energy 
following a power law (roughly $\sim E^{-2.8}$).
Again, the  behavior of the oscillation probability changes with the change 
in zones of $E-L$ plane. This has been  discussed in \cite{Samanta:2008af,Samanta:2008ag}. 
In case of proper binning of the data \cite{Samanta:2008af}, the precision of 
oscillation parameters improves. Then it helps in hierarchy discrimination
by reducing the effect of the ranges of 
the parameters, over which the marginalization is carried out.

In our previous work \cite{Samanta:2006sj}, the $\chi^2$ analysis has been carried out considering
the ratio of total up and total down going events for each resonance zone.  
Since the ratio up/down cancels all overall uncertainties, we considered only the energy 
dependent one. However, it should be noted here that the ratio of two Gaussian observables
is not an exact Gaussian function. Our previous $\chi^2$ study  assuming the up/down ratio as a Gaussian
function was motivated by the  cancellations of  all overall uncertainties.   
So, the result was an approximated one. 

It is  important to estimate the sensitivity of 
hierarchy determination with realistically measurable parameters of the experiments 
through a detailed analysis. 
Here, we have performed the $\chi^2$ analysis following Poissonian distribution and 
considering the number of events in the grids of the
$\log_{10} E - L^{0.4}$ plane of the muon. This follows Poissonian (Gaussian) distribution 
for a less (large) number of events. For a large number of events Poissonian distribution   
tends to a Gaussian one.
We have taken  into account 
 all possible systematic uncertainties using the pull method of the $\chi^2$ analysis.  
Expecting the lower bound of $\theta_{13}$ to be known from other experiments like 
Double Chooz \cite{Ardellier:2006mn} or NO$\nu$A \cite{Ayres:2004js}, 
we have also estimated the improved sensitivity considering the lower bound of $\theta_{13}$ for 
marginalization
as the input value of $\theta_{13}$ for generating the experimental data for the $\chi^2$ analysis.




\section{Oscillation formalism}
To understand the analytical solution of time evolution of neutrino propagation through matter,
 we adopt the 
so-called ``one mass scale dominance"  frame work: 
$|\Delta m_{21}^2| << | m_{3}^2 -m_{1,2}^2|$
\cite{Peres:2003wd,Akhmedov:2004ny}.

With this one mass scale dominance approximation, the survival probability of $\nu_\mu$
can be expressed as
\bea
P^{m}_{\mu \mu} &=&
{1 - \cos^2 \theta^m_{13} \; {\sin^2 2 \theta_{23}}} 
\nonumber\\
&&
\times\sin^2\left[1.27 \;\left(\frac{\da + A + \dam}{2}\right) \;\frac{L}{E} \right]
\nonumber \\
&& ~-~
\sin^2 \theta^m_{13}\; \sin^2 2 \theta_{23}
\nonumber \\
&& 
\times\sin^2\left[1.27 \;\left(\frac{\da + A - \dam}{2}\right) \;\frac{L}{E}
\right]
\nonumber \\
&& ~-~
{{\sin^4 \theta_{23}}} \;
{ { \sin^2 2\theta^m_{13} \;
\sin^2 \left[1.27\; \dam  \;\frac{L}{E} \right]
}}
\label{e:pmumu}
\eea

 The  mass squared difference ${{\dam}}$ and mixing angle
${ {\sin^22\theta_{13}^m}}$ in matter are related to their vacuum values by
\bea
\dam =
\nonumber
\sqrt{(\da \cos 2 \theta_{13} - A)^2 +
(\da \sin 2 \theta_{13})^2} 
\\
sin2\theta^m_{13}=
\frac{{\da \sin 2 \theta_{13}}}
{\sqrt{(\da \cos 2 \theta_{13} - A)^2 +(\da \sin 2 \theta_{13})^2.} }
\label{e:dm31}
\eea
where, the matter term 
$
A = 2 \sqrt{2} G_F n_e E = 7.63 \times 10^{-5}~{\rm eV}^2~\rho({\rm gm/cc})~
E({\rm GeV})~ \hbox{eV}^2.
$
\label{densm}
Here, $G_F$ and $n_e$ are the Fermi constant and the electron number density
in matter and $\rho$ is the matter density. The evolution equation for 
antineutrinos has the sign of $A$ reversed. 
 
From Eqs. \ref{e:pmumu} and \ref{e:dm31} it is seen that a
resonance in ${{P^{m}_{\mu \mu} }}$ will occur for 
neutrinos (antineutrinos) with normal hierarchy (inverted hierarchy) when
\be \sin^22\theta_{13}^m \rightarrow 1
 ~~~~{\rm or,}~~ A= \Delta m^2_{31} \cos2\theta_{13}. \ee

Then resonance energy can be expressed as 
\bea E = \left [\frac{1}{2\times 0.76 \times 10^{-4} Y_e} \right]
\left [\frac{| \Delta m_{31}^2|}{\rm eV^2} \cos2\theta_{13} \right ]
\left[\frac {\rm gm/cc}{\rho} \right ].
\label{e:resonance}\eea

%

\begin{figure*}[htb]
\includegraphics{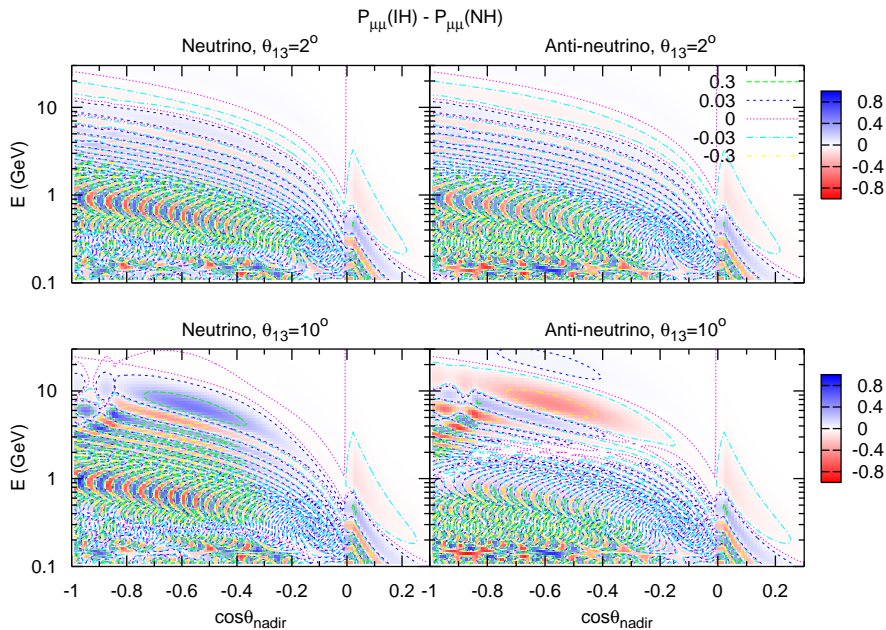}
\caption{\sf \small The oscillogram of the difference 
$P_{\mu\mu}(-\Delta m^{2}_{31}) - P_{\mu\mu}(\Delta m^{2}_{31})$ 
in the $E-\cos\theta_{\rm nadir}$ plane for $\theta_{13}=2^\circ$  (top) and 
$\theta_{13}=10^\circ$ (bottom) with 
$\nu_\mu$ (left) and $\bar\nu_\mu$ (right), respectively. We set 
$|\Delta m_{32}^2|=2.5\times 10^{-3}$eV$^2$, $\theta_{23}=45^\circ$, $\delta_{CP}=180^\circ$, 
and other oscillation parameters at their best-fit values.
}
\label{f:dpihnh}
\end{figure*}
The difference $\Delta P$ between $P_{\mu\mu}(\Delta m^{2}_{31})$ and $P_{\mu\mu}(-\Delta m^{2}_{31})$
has been plotted in Fig. \ref{f:dpihnh} for $\nu_\mu$ and $\bar\nu_\mu$ with $\theta_{13}=2^\circ$
and $\theta_{13}=10^\circ$, respectively, as an oscillogram in a two-dimensional plane of $E-\cos\theta^{\rm nadir}$.
We see that there exists a difference in all $E-\cos\theta^{\rm nadir}$ space. This is not due to the matter effect, but
due to the nonzero value of $\Delta m^{2}_{21}$. So this exists for $\theta_{13}=0^\circ$ also.
It is seen that the resonance zones are different for $\nu$ and $\bar\nu$ and both their areas and 
amplitudes squeeze with a decrease in $\theta_{13}$ values.

\section{The $\chi^2$ analysis}
The $\chi^2$ is calculated according to the Poisson probability distribution.
The binning the data is made  in  two-dimensional grids
in the plane of $\log_{10} E$ - $L^{0.4}$ of the muon.  The method for migration of number of events 
from neutrino to muon energy and zenith angle bins, the number of bins, the systematic 
uncertainties, and the cuts at the near horizons are described in \cite{Samanta:2008ag}.

The fact is that neutrino cross sections have not been precisely measured at all
energies, and the neutrino flux is also not precisely known  at every energy. There may arise
an energy dependent systematic uncertainty. We  may not always realize this. 
The program NUANCE considers difference processes of interactions at different energies
using the Monte Carlo method.
This may not be fully captured  when we generate theoretical data by folding the flux with total 
cross section
and smearing with the resolution functions\cite{Samanta:2008af}. The similar situation may happen
in real experiments also. The mass hierarchy is determined considering the difference in number of events 
between normal hierarchy (NH) and inverted hierarchy (IH). This arises for the resonance 
in  neutrino propagation through matter. It happens for some particular zones of energy and the baseline. 
So, there may arise a large difference
in estimated hierarchy sensitivities with and without consideration of this systematic 
uncertainty.
One can generate the experimental data for $\chi^2$ analysis in two ways: 
I) directly from NUANCE simulation for a given set of oscillation parameters with 1 Mton.year
exposure and then  binning the events in muon energy and zenith angle bins;
II) considering the oscillated atmospheric neutrino flux for a given set of parameters and then
folding with time of exposure, total cross section, detector mass and finally
smearing it with the energy-angle correlated resolution functions. This is similar with the
method of generating the theoretical data for $\chi^2$ analysis.

One can  generate the theoretical data directly from a huge data set, say, 
500 Mton.year data (to ensure the statistical error negligible) for {\bf each set} of oscillation
parameters and then reducing it to 1 Mton.year equivalent data from it, which would
be the more straightforward way for method I of generating experimental data. 
In this case, the effect of the above systematic uncertainty will not come into the play.  
The marginalization study with this method is  almost an undoable job in a  normal CPU. 
When we adopt different methods for generating theoretical and experimental data,
the significant effect of the above systematic uncertainties come into the results.
So, we adopt the same method for them and here we consider method II.

\begin{figure*}[htb]
\includegraphics[width=5.cm,angle=0]{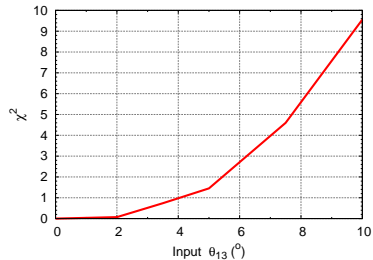}
\caption{\sf \small The variation of marginalized $\chi^2 {\rm(false)}$  
with different input values of $\theta_{13}$. 
}  
\label{f:chisexth}
\end{figure*}

\begin{figure*}[htb]
\includegraphics[width=5.cm,angle=0]{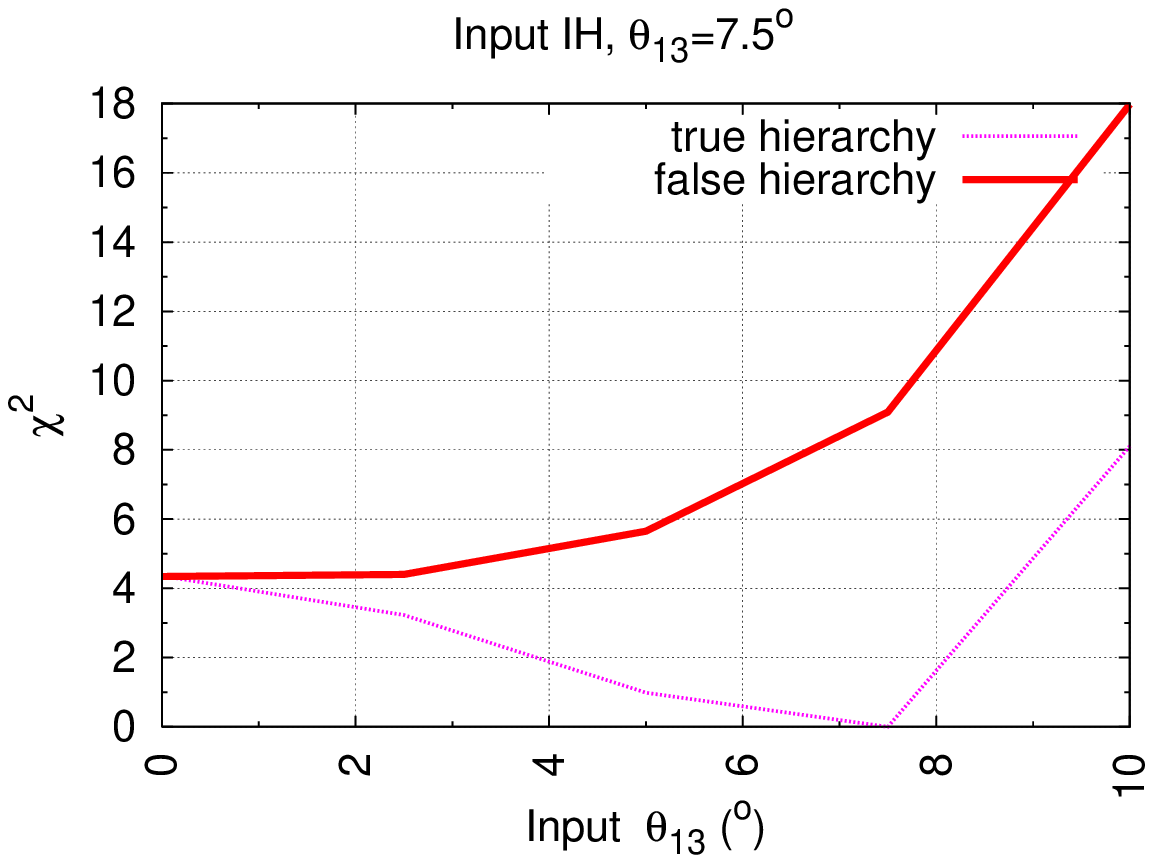}
\includegraphics[width=5.cm,angle=0]{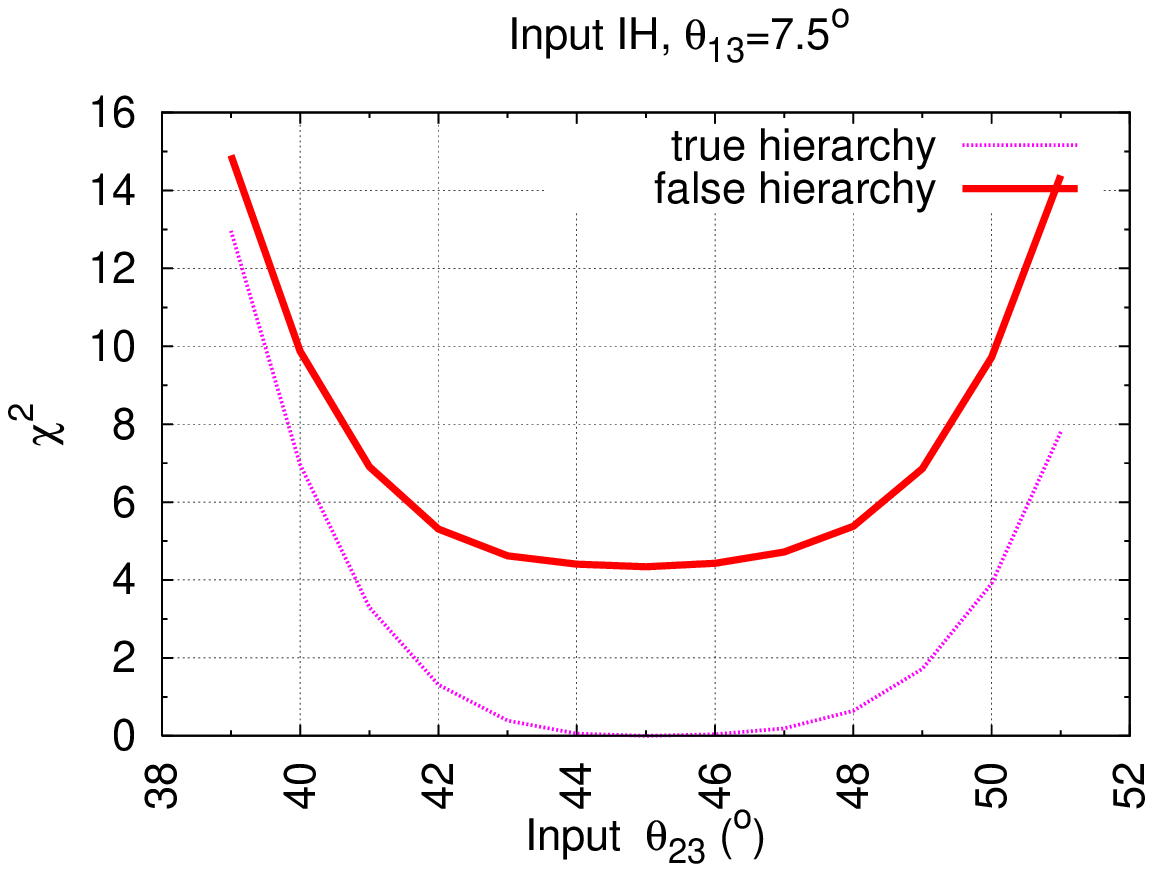}
\includegraphics[width=5.cm,angle=0]{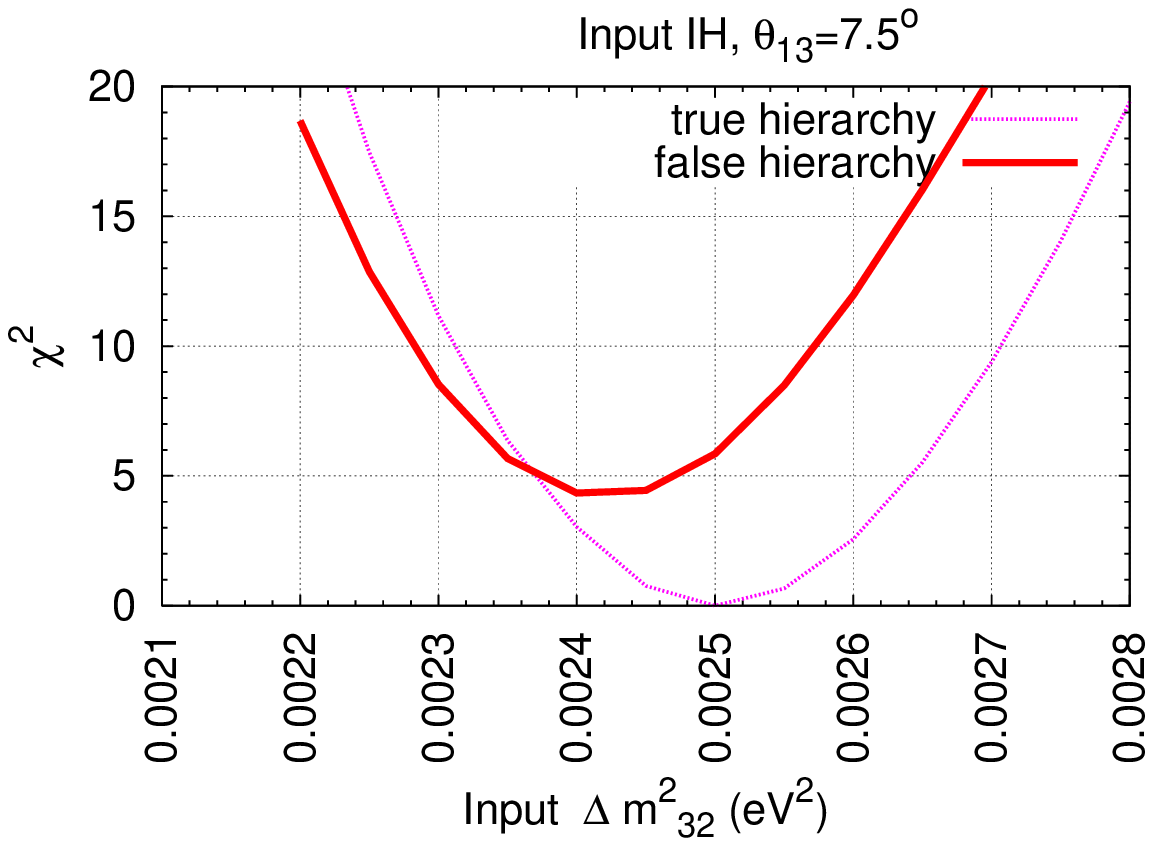}
\caption{\sf \small The variation of $\chi^2$ with $\theta_{13},~\theta_{23},$ and $\Delta m_{32}^2$,
respectively
for both true and false hierarchy. The $\chi^2$ is marginalized with respect to all oscillation 
parameters except with which it varies. 
}
\label{f:chionep}
\end{figure*}

\begin{figure*}[htb]
\includegraphics[width=6.cm,angle=0]{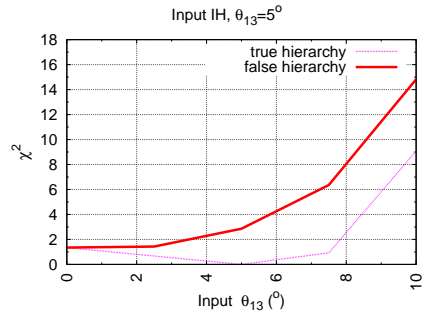}
\caption{\sf \small The same as Fig. \ref{f:chionep}, but with $\theta_{13}$ only and with input
$\theta_{13}=5^\circ$. 
}
\label{f:chith13_5}
\end{figure*}

\begin{figure*}[htb]
\includegraphics[width=6.cm,angle=0]{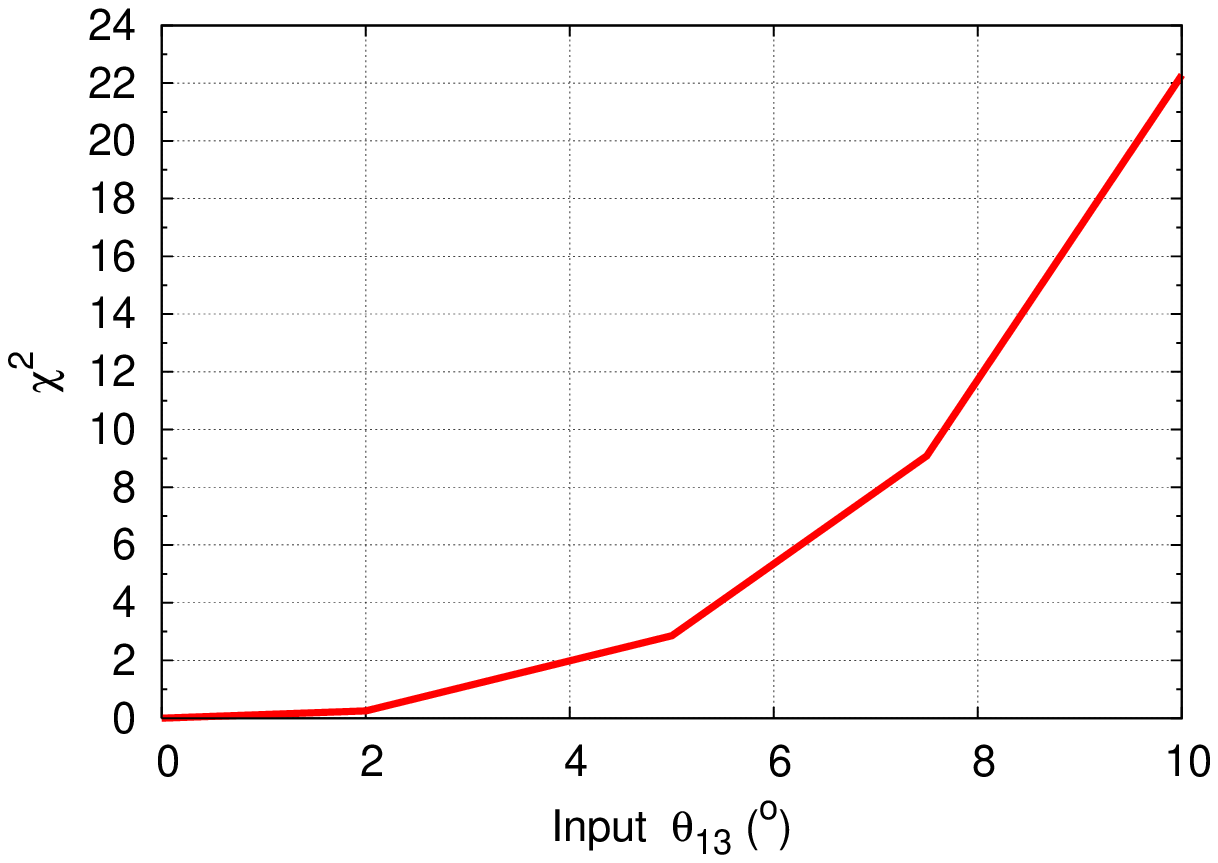}
\caption{\sf \small The same as Fig. \ref{f:chisexth}, but the lower limit of the 
range of $\theta_{13}$ for 
marginalization is the corresponding input value used for generating the experimental data set. 
This gives the lower limit of the sensitivity
corresponding to a lower limit of $\theta_{13}$ obtained from  other experiments.
}
\label{f:chilth13}
\end{figure*}

\section{Results}
We marginalize the $\chi^2$ over all the oscillation parameters
$\Delta m_{32}^2,~\theta_{23}, ~\theta_{13}$, and $\delta_{\rm CP}$
  along with solar oscillation
parameters $\Delta m_{21}^2$ and $\theta_{12}$ 
for both NH and IH
with $\nu$s and $\bar\nu$s separately for a given set of input data.
Then we find the total $\chi^2 [=\chi^2_\nu+\chi^2_{\bar\nu}]$.

We have chosen the range of $\Delta m_{32}^2=2.0-3.0\times 10^{-3}$eV$^2$,
$\theta_{23}=37^\circ - 54^\circ,$ $\theta_{13}=0^\circ - 12.5^\circ$ 
and $\delta_{\rm CP}=0^\circ - 360^\circ$. We set the range of 
$\Delta m_{21}^2=7.06-8.34\times 10^{-5}$eV$^2$ and $\theta_{12}=30.5^\circ - 40^\circ.$
However, the effect of $\Delta m_{21}^2$ comes in the
subleading order in the oscillation probability when $E\sim$ GeV and it is marginal.
%
We set the input of atmospheric oscillation  parameters
$|\Delta m^2_{32}|=2.5\times 10^{-3}$ eV$^2$, $\theta_{23}=45^\circ$, and $\delta_{\rm CP}=180^\circ$
and the solar parameters $\Delta m_{21}^2 = 7.9\times 10^{-5}$eV$^2$ and $\theta_{12}=33^\circ$.
We set IH as input.  The variation of marginalized $\chi^2$ for different input values of 
$\theta_{13}$ is shown in Fig. \ref{f:chisexth}.


To see the effect of marginalization over the ranges of parameters, we have shown  the 
variation of $\chi^2$ for both true and false hierarchy  as a function of $\theta_{13},~
\theta_{23}$ and $|\Delta m_{32}^2|$ in Fig. \ref{f:chionep}. Here,
 we have marginalized the $\chi^2$ over all oscillation parameters 
except one with which it varies.  We find that significant improvement will come 
if the lower bound of $\theta_{13}$ improves from zero. This is demonstrated in the 
first plot in Fig. \ref{f:chionep}.
We also see from this figure that there is no significant effect of other parameters on 
marginalization since they are 
well determined in this experiment. We find from Fig. \ref{f:chith13_5} that
if the lower limit is $5^\circ$, the mass hierarchy can be determined at a confidence level $> 90\%$.

The absolute bounds of each oscillation parameter can also be seen in Fig. \ref{f:chionep} when one 
sees the curve for its true hierarchy. It should be noted here that in the case of method II,
the best-fit values are same with their input values, which is not always the case in analysis using
method I (discussed in \cite{Samanta:2008af}). However, the precision of the parameters
is almost same here with method I (which is used in our previous work
\cite{Samanta:2008af}).

We have further studied for the cases when the lower limit of the range of 
$\theta_{13}$ for marginalization is the input value for the experimental data set. 
Considering this constraint we have plotted 
the variation of $\chi^2$ with input values of $\theta_{13}$ in Fig. \ref{f:chilth13}.
This will give the lower limit of the
sensitivity of the mass hierarchy in this detector when the lower limit of $\theta_{13}$ is 
known from other experiments like Double Chooz \cite{Ardellier:2006mn} or NO$\nu$A \cite{Ayres:2004js}. 
A significant improvement is observed comparing Figs. \ref{f:chisexth} and \ref{f:chilth13}. 
%

{\it Acknowledgments:}
This research has been supported by funds from Neutrino Physics projects
at HRI. The use of excellent cluster computational
facility installed by the funds of this project  is also gratefully acknowledged.
The  general cluster facility of HRI has also been used at the initial stages of the work.


\begin{thebibliography}{99}

\bibitem{theory}
B. Pontecorvo, J. Exp. Theor. Phys. {\bf 33}, 549 (1957); Sov. 
Phys. JETP, {\bf 6}, 429 (1958).

\bibitem{Fukuda:1998mi}
  Y.~Fukuda {\it et al.}  [Super-Kamiokande Collaboration],
  Phys.\ Rev.\ Lett.\  {\bf 81}, 1562 (1998)
  [arXiv:hep-ex/9807003].

\bibitem{Fogli:2008ig}
  G.~L.~Fogli {\it et al.},
  Phys.\ Rev.\  D {\bf 78}, 033010 (2008)
  [arXiv:0805.2517 [hep-ph]].



\bibitem{mixing}
Z. Maki, M. Nakagawa and S. Sakata, Prog. Theor. Phys. {\bf 28}, 870 
(1962).
B. Pontecorvo, JETP {\bf 6} (1958) 429;
{\bf 7} (1958) 172;



\bibitem{Indumathi:2004kd}
  D.~Indumathi and M.~V.~N.~Murthy,
  Phys.\ Rev.\  D {\bf 71}, 013001 (2005)
  [arXiv:hep-ph/0407336].

\bibitem{Gandhi:2007td}
  R.~Gandhi, P.~Ghoshal, S.~Goswami, P.~Mehta, S.~U.~Sankar and S.~Shalgar,
  Phys.\ Rev.\  D {\bf 76}, 073012 (2007)
  [arXiv:0707.1723 [hep-ph]].


\bibitem{Petcov:2005rv}
 S.~T.~Petcov and T.~Schwetz,
  Nucl.\ Phys.\  B {\bf 740}, 1 (2006)
 [arXiv:hep-ph/0511277].

\bibitem{Athar:2006yb}
 See ``India-based Neutrino Observatory: Project Report. Volume I,''
available at \begin{verbatim} http://www.imsc.res.in/~ino/\end{verbatim}.


\bibitem{Gandhi:2008zs}
  R.~Gandhi, P.~Ghoshal, S.~Goswami and S.~Uma Sankar,
  Phys.\ Rev.\  D {\bf 78}, 073001 (2008)
  [arXiv:0807.2759 [hep-ph]].

\bibitem{Samanta:2008af}
  A.~Samanta,
  Phys.\ Rev.\  D {\bf 80}, 113003 (2009)
  [arXiv:0812.4639 [hep-ph]].



\bibitem{Casper:2002sd}
  D.~Casper,
  Nucl.\ Phys.\ Proc.\ Suppl.\  {\bf 112}, 161 (2002)
  [arXiv:hep-ph/0208030].

\bibitem{geant}
http://geant4.web.cern.ch/geant4/

\bibitem{Samanta:2006sj}
  A.~Samanta,
  Phys.\ Lett.\  B {\bf 673}, 37 (2009)
  [arXiv:hep-ph/0610196].

\bibitem{Samanta:2008ag}
  A.~Samanta,
  Phys.\ Rev.\ D {\bf 79}, 053011 (2009)
  [arXiv:0812.4640 [hep-ph]].


\bibitem{Samanta:2009hd}
  A.~Samanta,
  Phys.\ Rev.\  D {\bf 80}, 073008 (2009)
  [arXiv:0907.3978 [hep-ph]].

\bibitem{Ardellier:2006mn}
  F.~Ardellier {\it et al.}  [Double Chooz Collaboration],
  arXiv:hep-ex/0606025.

\bibitem{Ayres:2004js}
  D.~S.~Ayres {\it et al.}  [NOvA Collaboration],
  arXiv:hep-ex/0503053.








\bibitem{Peres:2003wd}
  O.~L.~G.~Peres and A.~Y.~Smirnov,
  Nucl.\ Phys.\  B {\bf 680} (2004) 479
  [arXiv:hep-ph/0309312].

\bibitem{Akhmedov:2004ny}
  E.~K.~Akhmedov, R.~Johansson, M.~Lindner, T.~Ohlsson and T.~Schwetz,
  JHEP {\bf 0404} (2004) 078
  [arXiv:hep-ph/0402175].






\end{thebibliography}
\end{document}